\documentclass[12pt]{article}
\usepackage{amsfonts}
\usepackage{amsmath,amssymb}               	
\usepackage{amsbsy}
\usepackage{graphics}
\usepackage{hyperref}
\usepackage{epsfig}
\usepackage{color}

\input epsf.sty

\newlength{\TZ}
\newcommand{\BEQ}{\begin{equation}}     
\newcommand{\BEA}{\begin{eqnarray}}
\newcommand{\BD}{\begin{displaymath}}
\newcommand{\EEQ}{\end{equation}}       
\newcommand{\EEA}{\end{eqnarray}}
\newcommand{\ED}{\end{displaymath}}
\newcommand{\eps}{\varepsilon}          
\newcommand{\D}{{\rm d}}                
\newcommand{\II}{{\rm i}}               
\newcommand{\demi}{\frac{1}{2}}         
\newcommand{\wit}[1]{\widetilde{#1}}    
\newcommand{\wht}[1]{\widehat{#1}}      
\newcommand{\rar}{\rightarrow}          
\newcommand{\ket}[1]{\left|#1\right\rangle}  
\newcommand{\semi}{\ltimes}             

\renewcommand{\vec}[1]{\boldsymbol{#1}} 

\newcommand{\vekz}[2]
     {\mbox{${\begin{array}{c} #1  \\ #2 \end{array}}$}}
\newcommand{\matz}[4] 
     {\mbox{${\begin{array}{cc} #1 & #2 \\ #3 & #4 \end{array}}$}}
     
\catcode`\@=11
\def\numberbysection{\@addtoreset{equation}{section}
        \def\theequation{\thesection.\arabic{equation}}}
\numberbysection

\begin{document}

\begin{titlepage}

\vskip 1.5 cm
\begin{center}
{\LARGE \bf Logarithmic correlators or responses in non-relativistic analogues of conformal invariance}
\end{center}

\vskip 2.0 cm
\centerline{{\bf Malte Henkel}$^a$ and {\bf Shahin Rouhani}$^{b}$}
\vskip 0.5 cm
\centerline{$^a$Groupe de Physique Statistique,
D\'epartement de Physique de la Mati\`ere et des Mat\'eriaux,}
\centerline{Institut Jean Lamour (CNRS UMR 7198), 
Universit\'e de Lorraine Nancy,} 
\centerline{B.P. 70239, F -- 54506 Vand{\oe}uvre l\`es Nancy Cedex, France}
\centerline{$^b$ Physics Department, Sharif University of Technology, Tehran, PO Box 11165-9161, Iran}

\begin{abstract}
Recent developments on emergence of logarithmic 
terms in correlators or response functions of models which exhibit 
dynamical symmetries analogous to conformal invariance 
in not necessarily relativistic systems are reviewed. 
The main examples of these are logarithmic 
Schr\"odinger-invariance and logarithmic conformal Galilean invariance. 
Some applications of these ideas to statistical physics are described. 

\end{abstract}

\end{titlepage}

\setcounter{footnote}{0} 

\section{Introduction}

Dynamical symmetries have become an increasingly important 
tool for the analysis of widely
different physical systems. One particularly well-studied 
instance is represented by those systems
admitting conformal invariance, especially in two dimensions. 
Conformal invariance has always been one of the central ingredients in string theory. 
In statistical physics, conformal invariance arises in many situations, 
usually for sufficiently local interactions,
as a `natural' extension of scale-invariance \cite{Polyakov70}. 
Since in two dimensions, the associated Lie algebras are infinite-dimensional, 
$2D$ conformal invariance furnishes particularly
powerful methods for the analysis of such systems \cite{Belavin84}. 

The considerable recent interest in non-relativistic analogues of the conformal algebra 
is on one hand motivated by studies
of the AdS/CFT correspondence, 
in particular for topologically massive gravity~\cite{Son08, Hartnoll09, Bala08, Maldacena98, Gubser:1998, Witten98, Gray13}
(with applications to the physics of cold atoms \cite{Fuertes09}); 
on the other hand by studies in the non-equilibrium statistical physics of physical ageing; 
or else in relationship to strongly anisotropic critical phenomena at equilibrium 
as exemplified by Lifshitz multicritical points. 
Therefore, people have considered variants of conformal transformations, 
where first a 
`time'-variable $t$ is distinguished with respect to the `space' variables 
$\vec{r}$ and then a strongly 
an\-iso\-tro\-pic/dyna\-mical scaling is introduced by considering the dilatations 
(with $\lambda= \mbox{\rm cste.}$)
\BEQ
t\mapsto \lambda^z t \;\; , \;\;
\vec{r}\mapsto \lambda \vec{r}
\EEQ
such that the {\em dynamical exponent} $z$ 
describes the distinct behaviour of `time' with respect to `space'. 

Indeed, the list of known sets of admissible generators of space-time transformations, 
related to conformal transformations
and which include some kind of dilatations, and  
which close into a Lie algebra is a rather short one.
In $d+1$ space-time dimensions one has:
\begin{enumerate}
\item the {\em conformal algebra} $\mathfrak{conf}(d+1)$ itself, in $d+1$ dimensions, with $z=1$.
\item when considering a non-relativistic contraction 
\BEQ \label{eq1} 
t \longrightarrow t \;\; , \;\; \vec{r} \longrightarrow \vec{r}/c \;\; ; \;\; c\to \infty
\EEQ
one obtains the {\em conformal Galilean algebra} {\sc cga}$(d)$, 
apparently first identified in \cite{Havas78}, but independently
rediscovered in different contexts \cite{Henkel97,Negro97}. 
It is usually obtained, by a contraction,
as the non-relativistic limit of the
$(d+2)$-dimensional conformal algebra 
(itself obtained by a non-relativistic holographic construction)~\cite{Henkel03a,Martelli09,Bagchi09a,Bagchi09b,Leigh09,Leigh10,Jottar10}. 
There is a known infinite-dimensional extension for any spatial dimension $d\geq 1$ \cite{Cherniha10}. For $d=1$, it
can be constructed from a contraction of a
pair of commuting Virasoro algebras \cite{Henkel02,Henkel06b,Bagchi09b}. 
In most representations, one has $z=1$, but
representations with $z=2$ are also known \cite{Henkel06b}.
\item in $d=2$ space dimensions, there exists the {\em exotic conformal Galilean algebra} {\sc ecga},
which is the central extension  of the non-semi-simple {\sc cga}$(2)$ \cite{Lukierski06}.

Besides well-known linear equations invariant under {\sc ecga} \cite{Martelli09},
invariant non-linear equations have also been found \cite{Zhang10,Cherniha10}.
All known representations have a dynamical exponent is $z=1$.
See \cite{Horvathy10} for a recent review and the relationship with non-commutative mechanics.
\item The oldest known example of non-conformal space-time transformation is given by the
{\em Schr\"odinger algebra} $\mathfrak{sch}(d)$, and was found by Jacobi in 1842/43 \cite{Jacobi1843} and Lie in 1881 \cite{Lie1881}. 
The known representations give a dynamical exponent $z=2$. 

Although well-known to mathematicians, physicists re-discovered it as a symmetry of 
free non-relativistic particles several times around 1970,
including \cite{Kastrup68,Hagen72,Niederer72,Jackiw72}. 
Non-linear examples of Schr\"odinger-invariant equations include the
Navier-Stokes equation \cite{Ovsiannikov80,Hassaine00,ORaif01} or Burger's equation \cite{Niederer78,Ivash97}. 

The Schr\"odinger algebra $\mathfrak{sch}(d)\subset \mathfrak{conf}(d+2)$ \cite{Burdet73}, 
but earlier claims\footnote{The contraction procedure in \cite{Barut73} almost discovered \mbox{\sc cga}(d).} that the
Schr\"odinger algebra could be obtained by a contraction from the conformal algebra are incorrect \cite{Henkel03a}.

When classifying non-relativistic conformal Newton-Cartan space-times with a fixed dynamical exponent $z$,
the two non-trivial solutions are (i) the conformal Galilei algebra \mbox{\sc cga}$(d)$ for light-like geodesics and
(ii) the Schr\"odinger algebra $\mathfrak{sch}(d)$ for time-like geodesics \cite{Duval09}. Remarkably, these two solutions
also appeared in a search for local scale-transformations which admit the M\"obius-transformations 
$t\mapsto (\alpha t+\beta)/(\gamma t+\delta)$ in time \cite{Henkel02}. 
\item As we shall see below, the common sub-algebra $\mathfrak{age}(d)$ of both the conformal Galilean algebra \mbox{\sc cga}$(d)$ and
the Schr\"odinger algebra $\mathfrak{sch}(d)$ plays an important r\^ole in slow relaxation processes far from equilibrium and related 
to what is known in material science as `physical ageing'. Since physical ageing may be formally defined by its three properties
of (i) slow relaxations, (ii) breaking of time-translation-invariance and (iii) dynamical scaling, the Lie algebra $\mathfrak{age}(d)$
permits more general co-variant transformations. Especially, this can be cast in the form that a non-equilibrium scaling operator
should be characterised in terms of {\em two independent scaling dimensions}, denoted here $x$ and $\xi$ \cite{Picone04,Henkel06a}; 
rather than a single one
as found for the Lie algebras $\mathfrak{conf}(d)$, $\mathfrak{sch}(d)$ and \mbox{\sc cga}$(d)$, where $\xi=0$. This additional
freedom will be seen to be important in the construction of the logarithmic extension and for the phenomenological comparison
with specific models. 

Known representations of $\mathfrak{age}(d)$ in terms of local coordinates changes have either $z=2$ or $z=1$. However, when taking
a coset with respect to the underlying invariant differential equation, 
representations for any value of $z$ are known \cite{Henkel11,Stoimen12}. The correct geometrical interpretation of such non-local
transformations is still an open question. 
\item There exists for $d=1$ a closed algebra with $z=\frac{3}{2}$ \cite{Henkel02}.
It is not yet clear how this might
fit into the general scheme of \cite{Duval09}, since it does not contain the full conformal structure and furthermore
its generators contain {\em fractional} space derivatives.
\end{enumerate}

It is natural to wonder about the quantum realisations of these symmetries, 
their representations and correlation functions. Some attempts have been made at constructing 
NRCFT~\cite{Son08, Henkel97, Roger06, Henkel03a,Henkel06b,Bagchi09a,Bagchi09b,Alishahiha09, Martelli09, Giulini96,Duval09}.  
Many interesting features have been discovered and many questions remain. 
In this paper, we look at the question of whether 
logarithmic correlators may appear in NRCFT's analogous to their relativistic counter parts~\cite{Saleur:1992,Gurarie93}.
Logarithmic Conformal Field Theories (LCFT's) arise when the action of the dilatation generator $L_0$ 
on primary fields is not diagonal; this may happen in some ghost theories such as the $c=-2$ theory~\cite{Gaberdiel96}. 
Generically LCFT's are non unitary theories; 
however applications of LCFT's to some statistical models have been 
suggested~\cite{Saleur:1992, Watts96, Rahimi97b, Gurarie:1997,Mathieu:2001, Caux96}. 
Excellent reviews of LCFT can be found within this issue. 
We therefore concentrate exclusively on the appearance of 
logarithmic conformal field theories with non relativistic symmetries (NR-LCFT). 

Some examples taking from physical ageing will be used for tests and illustration. 

This paper is organised as follows. Section~2 describes logarithmic Schr\"odinger-invariance, its descendent states and
associated new invariant equations, as well as the derivation of two-point functions. In section~3, an extension towards a parabolic
sub-algebra of a higher-dimensional conformal algebra is described. Physically, co-variance under this parabolic aub-algebra
implies a causality condition for the $n$-point function; hence these are to be interpreted as {\em response functions}, rather than
as correlators. This is important for later applications in non-equilibrium statistical physics. In section~4, the logarithmic
extension of the conformal Galilean algebra are described, including also the so-called `exotic' central extension in $d=2$ dimensions. 
Section~5 recalls briefly the context of physical ageing far from equilibrium and then describes the new features which can arise 
from the more general representations of the ageing algebra. Section~6 illustrates to what extent these two-point functions actually
describe the non-equilibrium linear response of two paradigmatic models: (i) the $1D$ Kardar-Parisi-Zhang equation and (ii) $1D$ directed
percolation (Reggeon field theory). Section~7 gives our conclusions.

\section{Logarithmic Schr\"odinger-invariance}

\subsection{Lie algebra}

The {\em Schr\"odinger group} is defined by the following set of space-time transformations
\BEQ\label{eq:rr}
t \mapsto t' = \frac{\alpha t +\beta}{\gamma t + \delta} \;\; , \;\;
\vec{r} \mapsto \vec{r}' = \frac{{\cal R}\vec{r} + \vec{v} t + \vec{a}}{\gamma t +\delta} \;\; ; \;\; 
\alpha \delta - \beta \gamma =1
\EEQ
where ${\cal R}\in {\sl SO}(d)$ is a rotation matrix, $\vec{v},\vec{a}\in\mathbb{R}^d$ are vectors and 
$\alpha,\beta,\gamma,\delta$ are real numbers. 

When concentrating on the changes in the coordinates $(t,\vec{r})\in \mathbb{R}_+ \times \mathbb{R}^d$, 
one often uses the infinitesimal
generators in the form (with $\partial_i = \partial/\partial r_i$)
\BEA
P_i &=& \partial_i \;\; , \;\; H = -\partial_t \;\; , \;\; B_i = t\partial_i \nonumber \\
J_{ij} &=& -\left( x_i\partial_j - x_j\partial_i\right) \label{2.2} \\
D &=& -\left( 2 t\partial_t + r_i \partial_i \right) \;\; , \;\; K = - \left( t r_i \partial_i + t^2\partial_t\right) \nonumber
\EEA
which span the algebra $\mathfrak{sch}^{(0)}(d)$. 
Herein, the generators $P_i,H,B_i$, together with the $J_{ij}$, make up the Galilei sub-algebra 
(still without a non-relativistic mass). The two new generators are those of
dilatations ($D$) and of `special' Schr\"odinger transformations ($K$). 
Lie observed that these additional space-time transformations send
solutions of the free diffusion equation to other solutions \cite{Lie1881}, 
provided the solutions are also transformed by a further
`companion function' \cite{Niederer72}, see below. Furthermore, it is easy to see that the generators $H,D,K$ form a Lie algebra
$\mathfrak{sl}(2,\mathbb{R})$. Finally, one has a semi-direct sum structure 
$\mathfrak{sch}^{(0)}(d) \cong \left( \mathfrak{sl}(2,\mathbb{R}) \oplus \mathfrak{so}(d)\right) \semi \mathfrak{t}(2d)$, where
$\mathfrak{t}(2d)$ is the $2d$-dimensional Abelian algebra generated by space translations and Galilean boosts. 

One may therefore inquire whether $\mathfrak{sch}(d)$ 
can be extended to an infinite-dimensional Lie algebra, to be written as a semi-direct
sum with one of the terms being isomorphic to a Virasoro algebra $\mathfrak{vir}$. 
Indeed, this can be done \cite{Henkel94}. The
resulting algebra is by now called {\em Schr\"odinger-Virasoro algebra} 
$\mathfrak{sv}(d) := \left(\mathfrak{vir}\oplus  \mathfrak{so}(d)\right) \semi \mathfrak{hei}(d) 
= \left\langle X_n, Y_m^{(i)}, M_n, R_n^{(ij)}\right\rangle_{n\in\mathbb{Z},m\in\mathbb{Z}+\demi,i,j=1,\ldots,d}$. 
In $d$ spatial dimensions, the generators read \cite{Henkel02}
\BEA
X_n &=& -t^{n+1}\partial_t -\frac{n+1}{2} t^n r_i \partial_i 
- \frac{\cal M}{4} n(n+1) t^{n-1} \vec{r}^2 - \frac{x}{2} (n+1) t^n 
\nonumber \\
Y_m^{(i)}  &=& - t^{m+\demi} \partial_i - \left(m+\demi\right){\cal M} t^{m-\demi} r_i \nonumber \\
M_n        &=& - t^{n} {\cal M} \label{2.3} \\
R_n^{(ij)} &=& -t^n \left( r_i \partial_j - r_j \partial_i \right) \nonumber
\EEA
The {\em Schr\"odinger algebra} $\mathfrak{sch}(d):=\left\langle X_{\pm 1,0}, Y_{\pm\demi}^{(i)}, M_0, R_0^{(ij)}\right\rangle$ is the largest
finite-dimensional sub-algebra. 
If one sets $x=0$ and ${\cal M}=0$, then one has the following correspondence between the generators 
(\ref{2.2},\ref{2.3}) of $\mathfrak{sch}^{(0)}(d)$:
\BEQ
Y_{-\demi}^{(i)} = - P_i \;,\; Y_{\demi}^{(i)} = B_i \; , \; X_{-1} = H \; , \; X_0 = \demi D \; , \; X_1 = K
\EEQ
The non-vanishing commutators of the generators (\ref{2.3}) are readily obtained 
(with $n,n'\in\mathbb{Z}$ and $m,m'\in\mathbb{Z}+\demi$) 
\BEA
[ X_n , X_{n'} ] &=& (n-m) X_{n+n'} \nonumber \\ {}
[ X_n , Y_m^{(j)} ] &=& \left( \frac{n}{2} -m \right) Y_{n+m}^{(j)} 
\nonumber \\ {}
[ X_n , M_{n'} ] &=& -n' M_{n+n'} \nonumber \\ {}
[ X_n , R_{n'}^{(jk)} ] &=& - n' R_{n+n'}^{(jk)} \nonumber \\ {}
[ Y_m^{(i)} , Y_{m'}^{(j)} ] &=& \delta_{i,j} \, (m-m') M_{m+m'} 
\label{gl:5:svComm} \\ {}
[ R_n^{(ij)} , R_{n'}^{(kl)} ] &=& 
  \delta_{i,k} R_{n+n'}^{(jl)} + \delta_{j,l} R_{n+n'}^{(ik)} 
- \delta_{i,l} R_{n+n'}^{(jk)} - \delta_{j,k} R_{n+n'}^{(il)}  \nonumber \\ {}
[ R_n^{(ij)}, Y_m^{(k)} ] &=& \delta_{i,k} Y_{n+m}^{(j)}  - \delta_{j,k} Y_{n+m}^{(i)} \nonumber 
\EEA
The invariant free Schr\"odinger equation can be formally written as ${\cal S}\phi=0$, with the Schr\"odinger operator
\BEQ \label{2.6}
{\cal S} = 2 M_0 X_{-1} - \vec{Y}_{-1/2}\cdot \vec{Y}_{-1/2} 
= 2{\cal M}\partial_t - \vec{\nabla}_{\vec{r}}\cdot \vec{\nabla}_{\vec{r}}
\EEQ
The invariance is stated by $[{\cal S},{\cal X}]=0$ for (almost) all ${\cal X}\in\mathfrak{sch}(d)$, 
with the two exceptions $X_{0,1}$, where
$[{\cal S},X_0]=-{\cal S}$ and $[{\cal S},X_1] = 2t{\cal S} -(2x-1)M_0$. 

It should be stressed that the generators (\ref{2.3}) 
explicitly contain the information on the co-variant transformation of the
`wave function' $\phi$, which are described by (i) the scaling dimension $x$ and (ii) the non-relativistic (!) mass $\cal M$. 
In particular, if the scaling dimension of the wave function $x=x_{\phi}=\demi$, 
the transformations of $\mathfrak{sch}(d)$ send
a solution of ${\cal S}\phi=0$ onto another solution. 
For a non-vanishing mass $\cal M$, two generators $Y_m^{(i)}$ need no longer commute. 
In particular, the mass generator $M_0$ acts as a central charge in $\mathfrak{sch}(d)$, 
since $[Y_{\demi}^{(i)}, Y_{-\demi}^{(j)}]=\delta_{ij} M_0$ 
and hence one achieves a central extension of the abelian
algebra  to the Heisenberg algebra: 
$\mathfrak{t}(2d) \to \mathfrak{hei}(d):=\left\langle Y_{\pm\demi}^{(i)},M_0\right\rangle_{i=1,\ldots,d}$. 

Since $\mathfrak{sch}(d)$ is not semi-simple, its space-time representations are all projective, 
as made explicit by the non-derivative terms
related to $\cal M$ in the generators (\ref{2.3}). True representations can be constructed \cite{Giulini96}
by first considering $\cal M$ as an additional variable and then dualising with respect to it
\BEQ \label{2.7}
\phi(t,\vec{r}) = \frac{1}{\sqrt{2\pi\,}\,} \int_{\mathbb{R}} \!\D\zeta\: \exp\left(-\II {\cal M}\zeta\right) \wht{\phi}(\zeta,t,\vec{r})
\EEQ
The free Schr\"odinger equation becomes a Klein-Gordon equation in light-cone coordinates
\BEQ
\left( -2\partial_{\zeta}\partial_t + \partial_i \partial_i \right) \wht{\phi} =0
\EEQ
and the metric in this space reads 
\BEQ
\D s^2 = 2 \D\zeta\D t + \D\vec{r}\cdot\D\vec{r}
\EEQ
One readily rewrites the generators (\ref{2.3}) in this basis. One advantage of this formulation is that the natural
inclusion $\mathfrak{sch}(d)\subset\mathfrak{conf}(d+2)$ becomes obvious. In addition, this procedure suggests are further
extension of the $\mathfrak{sch}(d)$ to so-called {\em parabolic sub-algebras} of $\mathfrak{conf}(d+2)$. These extensions can be
used to derive causality conditions for co-variant $n$-point functions \cite{Henkel03a}. We shall describe this in section~3. 

The mathematical theory of the Schr\"odinger-Virasoro algebra has been studied in considerable detail \cite{Roger06,Unterberger12}. 

\subsection{Descendant states}

In principle, there are two ways to describe infinitesimal coordinate transformation and the co-variant transformation of scaling
operators $\phi$ under these.\footnote{We follow Cardy \cite{Cardy96} and refer to $\phi$ as {\em scaling operators}. 
`Scaling fields' would be their canonically conjugate fields.} 
The first one is to include not only the changes of the coordinates $(t,\vec{r})$ into the generators, but also
the terms describing the transformation of $\phi$ itself. This convention was applied in (\ref{2.3}). Checking the Jacobi identities
then automatically guarantees that the transformation of $\phi$ is consistent with the space-time transformation. 

In this section, we shall follow the alternative route. The Lie algebra generators only contain the direct changes of the coordinates and
one explicitly writes the transformation of the $\phi$. In the case of Schr\"odinger-symmetry, scaling operators are characterised
by their scaling dimension $x$ and their mass $\cal M$:
\BEQ
{} \left[ X_0, \phi\right] = \frac{x}{2} \phi \;\; , \;\;
{} \left[ M_0, \phi\right] = {\cal M} \phi
\EEQ
Raising (lowering) operators are $X_n$, $M_n$ and $Y_{m}^{(i)}$ with $n,m>0$ ($n,m<0$). Formally,
\BEA
{}\left[ X_0, \left[ X_n, \phi\right]\right] &=& \left( \frac{x}{2} -n\right) \left[ X_n, \phi\right] \;\; , \;\; \nonumber \\
{}\left[ X_0, \left[ Y_m^{(i)}, \phi\right]\right] &=& \left( \frac{x}{2} -m\right) \left[ Y_m^{(i)}, \phi\right] \;\; , \;\;  \\
{}\left[ X_0, \left[ M_n, \phi\right]\right] &=& \left( \frac{x}{2} -n\right) \left[ M_n, \phi\right] \;\; , \;\; \nonumber
\EEA
Since the central charge $M_0$ commutes with all operators, none of them can modify the value of $\cal M$. 

Descendent operators will be built from the primary ones. Algebraically, 
a Schr\"odinger-primary operator $\phi$ will be characterised by being annihilated by all lowering operators
\BEQ
{} \left[ X_n, \phi\right] = 0 \;\; , \;\;
{} \left[ Y_m^{(i)}, \phi\right] = 0 \;\; , \;\;
{} \left[ M_n, \phi\right] = 0
\EEQ
for all $n,m>0$. In addition, using the usual operator-state correspondence, one may represent operators by states $\ket{x,{\cal M}}$. 
Hence, for a state with dimension $x$ and mass $\cal M$, one has
\BEQ
X_0 \ket{x,{\cal M}} = \frac{x}{2} \ket{x,{\cal M}} \;\; , \;\;
M_0 \ket{x,{\cal M}} = {\cal M} \ket{x,{\cal M}} 
\EEQ
{}From now on, we shall also restrict to $d=1$ dimensions, and drop the corresponding index. 
Since the mass $\cal M$ cannot be modified by a raising operator, one may simplify the notation and write $\ket{x,{\cal M}}\to \ket{x}$. 
The effect of the raising operators is then, with $n,m<0$
\BEQ
X_{-n} \ket{x} \to \ket{x+n} \;\; , \;\; Y_{-m}\ket{x} \to \ket{x+m} \;\; , \;\; M_{-n}\ket{x} \to \ket{x+n}
\EEQ
The first excited state is thus $Y_{-\demi}\ket{x}$. The second level is obtained either by $\left(Y_{-\demi}\right)^2\ket{x}$ or
by $X_{-1}\ket{x}$. However, if $x=1/2$, these two states are not independent and one rather has the first null vector
\BEQ
\ket{\chi_2} = \left( Y_{-\demi} Y_{-\demi} -2{\cal M} X_{-1} \right)\ket{x}
\EEQ
such that $\ket{\chi_2}=0$ gives back the Schr\"odinger equation (\ref{2.6}). The next null state is found at level 3 for $x=\frac{11}{6}$,
namely
\BEQ
\ket{\chi_3}= \left( 3 X_{-1} Y_{-\demi} - 2 Y_{-\frac{3}{2}} + \frac{3}{2{\cal M}} Y_{-\demi}^3 \right) \ket{x} 
\EEQ
Then $\ket{\chi_3}=0$ leads to another scale-invariant equation, namely~\cite{Nakayama10}:
\BEQ
\left( 3 t^2 \partial_t \partial_r - 2\left( t \partial_r - r {\cal M}\right) - \frac{3 t^2}{2{\cal M}} \partial_r^3 \right)\phi(t,r)=0
\EEQ

\subsection{Two-point function: non-logarithmic case} 

Two-point functions of primary scaling operators can be found by considering the action of the generators (\ref{2.2})
\BEA
{}\left[ X_n, \phi(t,\vec{r})\right] &=& \left( t^{n+1}\partial_t +\frac{n+1}{2} t^n r_i \partial_i 
+ \frac{\cal M}{4} n(n+1) t^{n-1} \vec{r}^2 + \frac{x}{2} (n+1) t^n \right) \phi(t,\vec{r})
\nonumber \\
{} \left[ Y_m^{(i)}, \phi(t,\vec{r})\right]  &=& 
\left( t^{m+\demi} \partial_i + \left(m+\demi\right){\cal M} t^{m-\demi} r_i\right) \phi(t,\vec{r})  \\
\left[ M_n, \phi(t,\vec{r}) \right]        &=&  t^{n} {\cal M} \phi(t,\vec{r}) \nonumber 
\EEA
If we had used the generators (\ref{2.3}) instead, we would have simply required that $X_n \phi=0$ etc., with the same result. 

Since the representation of $\mathfrak{sch}(d)$ under study is projective, 
some extra care is required for the treatment of the extra phases.
Indeed, it is necessary to introduce a conjugate $\phi^*$ to the scaling operator $\phi$, such that $\phi^*$ should have the opposite
mass of $\phi$, or formally\footnote{In quantum  mechanics, when ${\cal M}=\II m$ is purely imaginary, this just becomes the 
complex conjugate of the wave function. However, for the diffusion equation, when ${\cal M}$ is real, one must define a `conjugate'
of the real-valued function $\phi(t,\vec{r})$ as the so-called `response field' $\wit{\phi}(t,\vec{r})$ and which can be introduced
through the Janssen-de Dominicis action in  non-equilibrium field-theory, see e.g. \cite{Picone04}.} 
\BEQ
\left[ M_0, \phi^*(t,\vec{r})\right] = -{\cal M} \phi^*(t,\vec{r})
\EEQ
Requiring the co-variance under $\mathfrak{sch}(1)=\langle X_{\pm 1,0}, Y_{\pm\demi}, M_0\rangle$, one now looks for a
two-point function of quasi-primary scaling operators
\BEQ
F = F(t_1,t_2;r_1,r_2) =\left\langle \phi_1(t_1, r_1) \phi_2^*(t_2,r_2)\right\rangle
\EEQ
If these are scalars under rotations, it is enough to consider the
$1D$ case, since any two spatial points $\vec{r}_1,\vec{r}_2$ can be brought to lie on a fixed line.
Following \cite{Henkel94,Henkel10}, space- and time-translation-invariance imply $F=F(t,\vec{r})$ with $t=t_1-t_2$ and 
$\vec{r}=\vec{r}_1-\vec{r}_2$. 
The requirement of Galilei-covariance leads to
\BEA
Y_{1/2} F &=& \left[ -t_1 \frac{\partial}{\partial r_1} - {\cal M}_1 r_1 - t_2 \frac{\partial}{\partial r_2} - (-{\cal M}_2) r_2 \right] F 
\nonumber \\
&=& \left[ \left( -t\partial_r -{\cal M}_1 r\right) - r_2 \left(
{\cal M}_1 -{\cal M}_2\right) \right] F \:=\: 0
\EEA
This is only consistent with spatial translation-invariance if both terms in the second line vanish separately. Hence 
\BEA
\left( -t\partial_r -{\cal M}_1 r\right) F &=& 0 \label{gl:6:Y12}\\
{\cal M}_1 -{\cal M}_2 &=& 0 \label{gl:6:Barg}
\EEA
where the first one fixes the scaling function and the second one relates the
two `masses' and is an example of the well-known {\em Bargman superselection rule}s \cite{Bargman54}. Next, combining
dilatation-invariance with the translation-invariances gives 
\BEQ \label{gl:6:X0}
X_0 F = \left[ -t\partial_t -\frac{1}{2} r\partial_r -\frac{1}{2}(x_1+x_2)\right] F = 0
\EEQ
and finally co-variance under the special transformation gives
\BEA
X_1 F &=& \left[ -t_1^2\frac{\partial}{\partial t_1} -t_2^2\frac{\partial}{\partial t_2} 
- t_1 r_1 \frac{\partial}{\partial r_1} - t_2 r_2 \frac{\partial}{\partial r_2}
-\frac{{\cal M}_1}{2} r_1^2  +\frac{{\cal M}_2}{2} r_2^2 -x_1 t_1 -x_2 t_2
\right] F
\nonumber \\
&=& \left[ -t^2\partial_t -tr\partial_r -\frac{{\cal M}_1}{2} r^2-x_1 t
\right] F(t,{r}) =  0
\label{gl:6:X1}
\EEA
where both dilatation-invariance as well as both consequences of Galilei-invariance were used. 
In order to find $F$, multiply eq.~(\ref{gl:6:X0}) by $-t$ and add to eq.~(\ref{gl:6:X1}) 
and then multiply eq.~(\ref{gl:6:Y12}) with $-r/2$ and also add. The result is the condition
\BEQ
t r (x_1 - x_2) F(t,r)=0
\EEQ
which implies that $x_1=x_2$. Using this condition, the solution of the
remaining system (\ref{gl:6:Y12},\ref{gl:6:X0}) is elementary and gives \cite{Henkel94}, where
$f_0$ is a normalisation\index{two-point function}\index{Schr\"odinger-invariance} constant,
\BEQ \label{gl:6:Schr2P}
\left\langle \phi_1(t_1,\vec{r}_1)\phi_2^*(t_2,\vec{r}_2)\right\rangle =\delta_{x_1,x_2}\,
\delta_{{\cal M}_1,{\cal M}_2}\, f_0\, (t_1-t_2)^{-x_1}
\exp\left[-\frac{{\cal M}_1}{2}\frac{(\vec{r}_1-\vec{r}_2)^2}{t_1-t_2}\right]
\EEQ
This is essentially the heat-kernel solution (Green's function) of the diffusion equation.
Our implicit physical convention assumes  ${\cal M}_1>0$. 

Several aspects of this result quite closely resemble the conformally invariant two-point function, especially
the constraint $x_1=x_2$ on the scaling dimensions. 

\subsection{Two-point function: logarithmic case} 

The analogy of (\ref{gl:6:Schr2P}) with conformal invariance suggests that a logarithmic form might be found
by assuming a logarithmic structure for the quasi-primary operators. This means that the scaling dimensions should
be taken in a Jordan form (we restrict to the most simple case of rank 2). Hence there is a pair $(\phi,\psi)$ of primary operators
which form a reducible, but indecomposable representation
\BEA
X_0 \phi(z) \ket{0} &=& \frac{x}{2} \phi(z) \ket{0} \nonumber \\
X_0 \psi(z) \ket{0} &=& \frac{x}{2} \psi(z) \ket{0} + \phi(z) \ket{0} 
\EEA
Two-point functions are now to be formed from the operators $\phi$ and $\psi$. By the same procedure as in the above subsection, we find
a set of coupled differential equations for the three possible two-point functions, with the solutions
\BEA
\left\langle \phi_1(t_1,r_1) \phi_2^*(t_2,r_2)\right\rangle &=& 0 \nonumber \\
\left\langle \phi_1(t_1,r_1) \psi_2^*(t_2,r_2)\right\rangle &=& 
  \delta_{x_1,x_2} \delta_{{\cal  M}_1,{\cal M}_2} t^{-x_1} \exp\left[-\frac{{\cal M}_1}{2} \frac{r^2}{t}\right] b \label{2.29} \\
\left\langle \psi_1(t_1,r_1) \psi_2^*(t_2,r_2)\right\rangle &=& 
  \delta_{x_1,x_2} \delta_{{\cal  M}_1,{\cal M}_2} t^{-x_1} \exp\left[-\frac{{\cal M}_1}{2} \frac{r^2}{t}\right] 
  \left( c - b \ln t\right) \nonumber 
\EEA
with $t=t_1-t_2$ and $r=r_1-r_2$ and where $b,c$ are  free normalisation constants.  

Alternatively, one may also work with nilpotent variables
\BEQ
\theta^2 = 0 
\EEQ
We then assume the existence of quasi-primary operators and states \cite{Moghimi00}
\BEA
\Phi(z,\theta) &=& \phi(z) + \theta \psi(z) \nonumber \\
\Phi(z,\theta) \ket{0} &=& \ket{x/2 + \theta} \\
X_0 \ket{x/2 + \theta} &=& (x/2+\theta) \ket{x/2+\theta} \nonumber
\EEA
and we define the two-point function as
\BEQ \label{2.32}
\left\langle \Phi_1( t_1, r_1, \theta_1)\Phi_2^*( t_2, r_2, \bar{\theta}_2)\right\rangle = F(t_1,t_2;r_1,r_2;\theta_1,\bar{\theta}_2)
\EEQ 
and a conjugate nilpotent variable $\bar{\theta}_2$ appeared in the conjugate operator $\Phi_2^*$. Requiring co-variance under the
Schr\"odinger algebra, we find
\BEQ
F = \delta_{x_1,x_2} \delta_{{\cal  M}_1,{\cal M}_2} t^{-x_1-\theta_1-\bar{\theta}_2} 
\exp\left[-\frac{{\cal M}_1}{2} \frac{r^2}{t}\right] \left( b \left(\theta_1 + \bar{\theta}_2\right) + c \theta_1\bar{\theta}_2\right)
\EEQ
and after expanding,  leads to
\BEQ \label{2.34}
F = \delta_{x_1,x_2} \delta_{{\cal  M}_1,{\cal M}_2} t^{-x_1} 
\exp\left[-\frac{{\cal M}_1}{2} \frac{r^2}{t}\right]
\left( b \left(\theta_1 + \bar{\theta}_2\right) + \theta_1\bar{\theta}_2\left( c - 2b \ln t \right) \right)
\EEQ
However, we can expand the two-point function (\ref{2.32}) as follows: 
\BEA
\lefteqn{\left\langle \Phi_1( t_1, r_1, \theta_1)\Phi_2^*( t_2, r_2, \bar{\theta}_2)\right\rangle}  \nonumber \\
&=& \left\langle \phi_1( t_1, r_1)\phi_2^*( t_2, r_2)\right\rangle 
+ \bar{\theta}_2 \left\langle \phi_1( t_1, r_1)\psi_2^*( t_2, r_2)\right\rangle
\nonumber \\
& & + {\theta}_1 \left\langle \psi_1( t_1, r_1)\phi_2^*( t_2, r_2)\right\rangle
+ \theta_1 \bar{\theta}_2 \left\langle \psi_1( t_1, r_1)\psi_2^*( t_2, r_2)\right\rangle
\EEA
Comparison with the expansion of the two-point function in (\ref{2.34}) reproduces all three two-point functions in (\ref{2.29}). 

\section{Extension to parabolic sub-algebras and implications for causality}
 
There is a natural extension of the Schr\"odinger algebra which allows to derive  causality properties
of the co-variant $n$-point functions from purely algebraic criteria. Recall the root diagramme associated with
a Lie algebra \cite{Knapp86}, for the special case $d=1$: to each generator ${\cal X}\in\mathfrak{sch}(1)$
one associates a planar vector $\stackrel{\to}{\vec{x}}\in\Delta$ on a root lattice. 
Under this correspondence, forming the commutator
$[{\cal X},{\cal X}']={\cal Y}$ corresponds to vector addition 
$\stackrel{\to}{\vec{x}} + \stackrel{\to}{\vec{x}'}=\stackrel{\to}{\vec{y}}$. 
If that vector sum $\stackrel{\to}{\vec{y}}$ falls outside the lattice $\Delta$, 
it is understood that ${\cal Y}=0$.

\begin{figure}[tb]
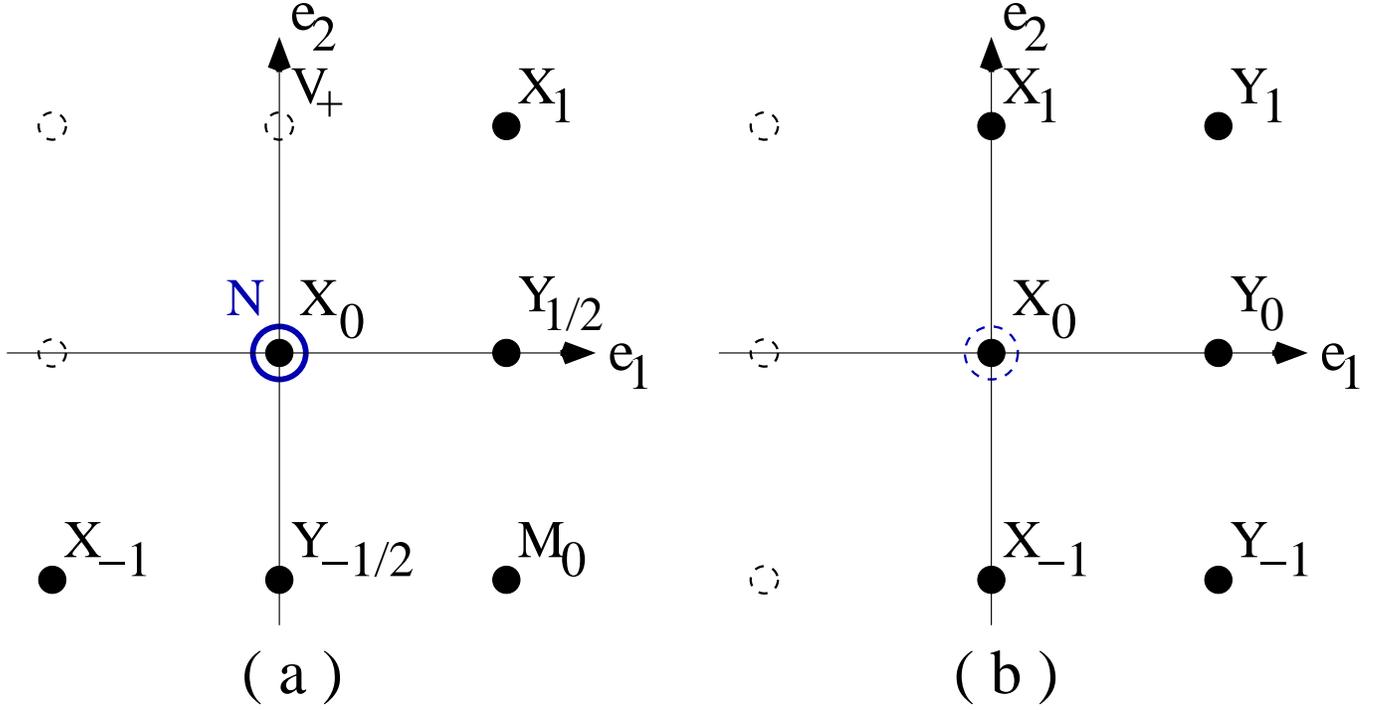

\centerline{\psfig{figure=log_sch_age_fig1a.eps,width=3.4in,clip=} ~~~~
\psfig{figure=log_sch_age_fig1b.eps,width=3.4in,clip=}}
\caption[figab]{\label{fig1ab} Root diagrammes of some 
sub-algebras of the complex Lie algebra $B_2$. The roots
of $B_2$ are indicated by the full and broken dots, 
those of the sub-algebras by the full dots only. \\
(a) Schr\"odinger algebra 
$\mathfrak{sch}(1)=\left\langle X_{\pm 1,0}, Y_{\pm 1/2}, M_0\right\rangle$ and the
maximal parabolic sub-algebra 
$\wit{\mathfrak{sch}}(1)=\mathfrak{sch}(1)+\mathbb{C} N$. \\
(b) Conformal Galilean algebra 
$\mbox{\sc cga}(1) = \left\langle X_{\pm 1,0}, Y_{\pm 1,0}\right\rangle$. 
}
\end{figure}

In figure~\ref{fig1ab}a, this is illustrated for the Lie algebra $\mathfrak{sch}(1)$. Since it closes as a Lie algebra,
the set of associated points must be convex. For example, it can be readily seen that the generator $M_0$ indeed is central. 
Furthermore, the same diagramme also illustrates the inclusion $\mathfrak{sch}(1)\subset \mathfrak{conf}(3) \cong B_2$, one of
the well-known simple Lie algebras of rank 2 in the Cartan classification and isomorphic to the algebra of conformal transformations
in 3 dimensions.  

\begin{figure}[tb]
\centerline{\psfig{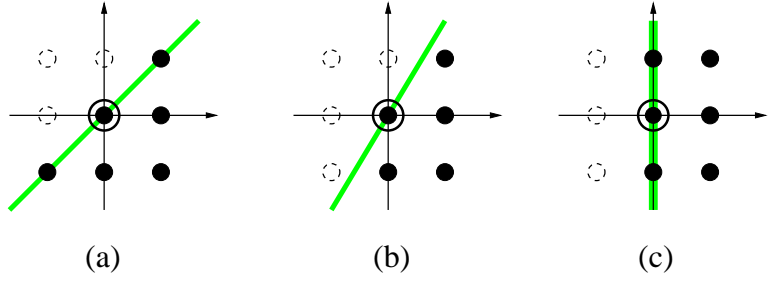}}
\caption[fig2]{\label{fig2} Classification of the parabolic sub-algebras of the simple complex algebra $B_2$. 
The generators are indicated by the full dots and the double-degenerate Cartan sub-algebra $\mathfrak{h}$. 
The three non-isomorphic parabolic sub-algebras of $B_2$ are 
(a) the extended Schr\"odinger algebra $\wit{\mathfrak{sch}}(1)$ (b) the extended ageing algebra
$\wit{\mathfrak{age}}(1)$ (c) the extended conformal Galilean algebra $\wit{\mbox{\sc cga}}(1)$. 
}
\end{figure}

There is an intermediate step between the algebra $\mathfrak{sch}(1)$ and the full conformal algebra $\mathfrak{conf}(3)$. 
These are the {\em parabolic sub-algebras} (in this case of $B_2$). By definition \cite{Knapp86}, a parabolic sub-algebra consists
of the Cartan sub-algebra $\mathfrak{h}$ and the set of all `positive' roots. A root is called positive, if it is to the right of a
straight line which passes through the origin of the graph, see figure~\ref{fig2}. 
For $\mathfrak{sch}(1)$ this straight line rune with a slope of $45^{\circ}$
through the root diagramme of $B_2$, see figure~\ref{fig1ab}a. With respect to $\mathfrak{sch}(1)$, the parabolic sub-algebra
$\wit{\mathfrak{sch}}(1) := \mathfrak{sch}(1) + \mathbb{C} N$ contains an extra generator $N$. Several set of roots can be mapped onto
each other by elements of the Weyl group and such pairs of sets are isomorphic as Lie algebras. Because of the Weyl symmetries,
the slope $\sigma$ of the straight line can be taken to lie between 
$1$ and $\infty$. In figure~\ref{fig2}, it is illustrated that there are three
non-isomorphic parabolic sub-algebras of $B_2\cong \mathfrak{conf}(3)$. 
The generic case, with $1<\sigma<\infty$, is the extended ageing algebra 
$\wit{\mathfrak{age}}(1) = \mathfrak{age}(1) + \mathbb{C} N$, which we shall discuss below in section~5. If the slope $\sigma=1$,
one has the extended Schr\"odinger algebra $\wit{\mathfrak{sch}}(1)$ and for a slope $\sigma=\infty$ one has the
extended conformal Galilei algebra $\wit{\mbox{\sc cga}}(1)$. For a formal proof of this classification, see \cite{Henkel03a}.  

This extension to $\wit{\mathfrak{sch}}(1)$ is more easy to see in the dual variables introduced in (\ref{2.7}). 
The representation (\ref{2.3}) is rewritten as 
\BEA
X_n &=& \frac{\II}{2}(n+1)n t^{n-1} {r}^2\partial_{\zeta}
-t^{n+1}\partial_t - \frac{n+1}{2}t^n {r}\partial_r  
- \frac{n+1}{2} x t^n \nonumber \\
Y_m &=& \II \left( m + \demi\right) t^{m-1/2} r\partial_{\zeta} - t^{m+1/2} \partial_j 
\nonumber \\
M_n &=&  \II t^n \partial_{\zeta}  \label{3.2} 
\EEA
The extension to the maximal parabolic sub-algebra\footnote{In the sense that 
adding any further generator brings one back to the full algebra $B_2$.} is achieved by including the
generator \cite{Henkel03a}
\BEQ \label{3.3}
N := \zeta\partial_{\zeta} - t \partial_t + \xi\,.
\EEQ
It is well-known that co-variance under this extra generator is sufficient to derive causality conditions of the form $t>0$
for the two-point functions, and also similarly for the three-point function \cite{Henkel03a}. 

We wish to study the consequences for logarithmic representations, 
built in analogy with those of the Schr\"odinger algebra $\mathfrak{sch}(1)$. Formally, this will again be achieved
by replacing the scaling dimension $x$ by a Jordan matrix $x\mapsto \left(\begin{array}{ll} x & 1 \\ 0 & x \end{array}\right)$. 
The co-variant two-point functions, built from quasi-primary scaling operators 
$\left(\vekz{\phi_i}{\psi_i}\right)$, are 
\BEA
\wht{F}(\zeta,t,r) &:=& \left\langle 
\wht{\phi}_1(\zeta_1,t_1,r_1)\wht{\phi}_2(\zeta_2,t_2,r_2)\right\rangle \nonumber \\
\wht{G}_{12}(\zeta,t,r) &:=& \left\langle 
\wht{\phi}_1(\zeta_1,t_1,r_1)\wht{\psi}_2(\zeta_2,t_2,r_2)\right\rangle \nonumber \\
\wht{G}_{21}(\zeta,t,r) &:=& \left\langle 
\wht{\psi}_1(\zeta_1,t_1,r_1)\wht{\phi}_2(\zeta_2,t_2,r_2)\right\rangle \label{3.4} \\
\wht{H}(\zeta,t,r) &:=& \left\langle 
\wht{\psi}_1(\zeta_1,t_1,r_1)\wht{\psi}_2(\zeta_2,t_2,r_2)\right\rangle \nonumber
\EEA
where $\zeta=\zeta_1-\zeta_2$, $t=t_1-t_2$ and $r=r_1-r_2$ and the three translation-symmetries are taken into account.  
As in logarithmic conformal invariance and analogously to the calculations in section~2, one may derive a set of linear
first-order differential equations for these four two-point functions. For $\mathfrak{sch}(1)$-covariance alone, 
the result \cite{Henkel12a} is, subject to the additional constraint $x := x_1 = x_2$,
that $\wht{F}=0$ and
\BEA
\wht{G}_{12}=\wht{G}_{21} = \wht{G}(t,u) &=& |t|^{-x} \wht{g}\left( u |t|^{-1}\right) \nonumber \\
\wht{H}(t,u) &=& |t|^{-x} \left( \wht{h}\left( u |t|^{-1}\right)-\ln|t| \wht{g}\left( u |t|^{-1}\right)\right)
\label{3.5}
\EEA
where $u=2\zeta t + \II r^2$ and $\wht{g}$ and $\wht{h}$ are arbitrary (differentiable) functions. 
Backtransforming to the masses ${\cal M}_{1,2}$, one reproduces the known form (\ref{2.29}). 

New results can be found by requiring co-variance under the larger algebra $\wit{\mathfrak{sch}}(1)$. For the logarithmic case, $\xi$
should be replaced by a matrix. Furthermore, it can be shown that this matrix also must be of Jordan form \cite{Henkel12a}
\BEQ 
N := \zeta\partial_{\zeta} - t \partial_t + \left( \begin{array}{ll} \xi & \xi' \\ 0 & \xi\end{array}\right).
\EEQ
Co-variance under $N$ fixes the two undetermined scaling functions in (\ref{3.5}), with the result
\BEA
\wht{G}(\zeta,t,r) &=& \wht{g}_0\, |t|^{-x}\: 
\left( \frac{2\zeta t+\II r^2}{|t|}\right)^{-x-\xi_1-\xi_2} \nonumber \\
\wht{H}(\zeta,t,r) &=& |t|^{-x}\: \left( \frac{2\zeta t+\II r^2}{|t|}\right)^{-x-\xi_1-\xi_2} \\
& & \times \left( \wht{h}_0 + \wht{g}_0 (1+\xi_1'+\xi_2') 
\ln\left(\frac{2\zeta t+\II r^2}{|t|}\right) - \wht{g}_0\,\ln|t|\right) \nonumber 
\EEA
where $\wht{g}_0$ and $\wht{h}_0$ are normalisation constants. One may transform this back to the masses ${\cal M}_{1,2}>0$. 
Again, one recovers exactly the previously found forms (\ref{2.29}), but now with the important extra information, that
$t=t_1-t_2>0$. If that condition is not met, the two-point function vanishes \cite{Henkel12a}. 

Causality conditions of this kind suggest that the two-point functions 
just calculated are better not interpreted as two-time correlators
$C(t,s)=\langle\phi(t)\phi(s)\rangle=C(s,t)$, but 
rather as linear response functions $R(t,s) 
= \left.\frac{\delta \langle\phi(t)\rangle}{\delta h(s)}\right|_{h=0} 
= \Theta(t-s) \wit{r}(t,s)$,
which measures the response of an average $\langle \phi(t)\rangle$ 
at some time $t$ with respect to an external perturbation which naturally should have occurred at an {\em earlier} time $s<t$,
as expressed by the Heaviside function $\Theta(t-s)$.

\section{Logarithmic conformal Galilean algebra, including the exotic case}

\subsection{ conformal Galilean algebra}

The  conformal Galilean algebra (\mbox{\sc cga}) can be obtained directly by 
contraction from the conformal algebra~\cite{Havas78}. 
Alternatively, one may also start from the so-called $l$-Galilei algebra \cite{Henkel97} and recognise the \mbox{\sc cga} as
the $l=1$ special case \cite{Henkel97, Negro97}. The embedding $\mbox{\sc cga}(1)\subset \mathfrak{conf}(3) \cong B_2$ and the
associated parabolic extension is illustrated in figure~\ref{fig2}c. 
This algebra is a straightforward generalisation of the transformations defined by eq.~\eqref{eq:rr}. 
We admit here a more general from;
\begin{equation}
t \mapsto t' = \frac{\alpha t + \beta }{\gamma t + \delta} \;\; , \;\;
\vec{r} \mapsto \vec{r'} =  \frac{\mathcal{R}\vec{r} + t^{2l}\vec{b}_{2l} + \cdots + \vec{b}_1t + \vec{b}_0}{\gamma t + \delta} \;\; , \;\;
\alpha \delta - \beta \gamma = 1.
\end{equation}
The algebra of the symmetry operators closes only for $l\in\demi\mathbb{Z}$. 
Recall that the dynamical exponent $z=1/l$ is related to the inverse of $l$,
\begin{equation}
t \mapsto \lambda^2 t, \quad \vec{r} \mapsto \lambda^{2l} \vec{r}
\end{equation}
thus only certain non-relativistic systems are included in this scheme. 
The case of Schr\"odinger symmetry corresponds to $l=1/2$. The case of $ l=1 $ leads to the \mbox{\sc cga}. 
Clearly, the dynamical exponent associated with this representation of the \mbox{\sc cga} is $z=1$. 
In $ d+1 $-dimensions, in addition to the usual generators of the Galilean algebra, 
$\{J_{i,j},H,P_i,B_i\}$ in eq.~(\ref{2.2}), $\mbox{\sc cga}(d)$ has $ d+2 $ more generators:
\begin{equation} \label{cga4.3}
D = -(t \partial_t + r_i \partial_i), \qquad K = -(2tr_i\partial_i + t^2 \partial_t), \qquad K_i = t^2 \partial_i.
\end{equation}
In contrast to the projective representations of the Schr\"odinger algebra, there is no analogue of a non-relativistic `mass' $\cal M$. 
Similar to the Schr\"odinger algebra, the $\mbox{\sc cga}(d)$ admits an affine extension \cite{Henkel02}, 
often called the {\it full \mbox{\sc cga}}:
As in previous sections, one may include into the generators immediately also the terms which describe the
co-variant transformation of the scaling operators. Then the generators of the full \mbox{\sc cga} 
may be written as follows \cite{Cherniha10}
\BEA X_n &=&
- t^{n+1}\partial_t - (n+1) t^n \vec{r}\cdot\vec{\nabla}_{\vec{r}}
- n(n+1) t^{n-1} \vec{\gamma}\cdot\vec{r} - x (n+1)t^n
\nonumber \\
Y_n^{(j)} &=& - t^{n+1} \partial_{j} - (n+1) t^n \gamma_j  \label{cga4.4} \\
J_n^{(jk)} &=& - t^n \bigl( r_j \partial_{k} -  r_k \partial_{j} \bigr)
- t^n \bigl( \gamma_j \partial_{\gamma_k}-\gamma_k
\partial_{\gamma_j}\bigr) \nonumber 
\EEA
where $\vec{\gamma}=(\gamma_1,\ldots,\gamma_d)$ 
is a vector of dimensionful constants, $x$ is again a scaling dimension 
%
and $n\in\mathbb{Z}$. The maximal finite-dimensional sub-algebra is the \mbox{\sc cga}(d), 
see figure~\ref{fig1ab}b for a root diagramme for $d=1$. If one sets $x=0$ and $\vec{\gamma}=\vec{0}$, the correspondence with
(\ref{cga4.3}) reads:
\BEA
X_{-1} &=& H \;\; , \;\; X_0  \:=\: D \;\; , \;\;  X_{1}  \:=\: K,  \nonumber \\
Y_{-1}^{(i)} &=& -P_i \;\; , \;\;  Y_0^{(i)} \:=\: -B_i \;\; , \;\;  Y_1^{(i)} \:=\: -K_i.
\EEA
The commutation relations of the full \mbox{\sc cga} are (with the habitual correspondence $J_{ij}=-J_{ji} \leftrightarrow J^{(a)}$):
\begin{equation}
\begin{aligned}
 \lbrack X_n, X_m  \rbrack  &= (m-n) X_{n+m},  \quad &  \lbrack X_m , J_n^{(a)} \rbrack & = -n J_{n+m}^{(a)}, \\
 \lbrack J_m^{(a)}, J_n^{(b)} \rbrack &= -{f^{ab}}_{c} J_{n+m}^{(c)}, \quad &  \lbrack X_m, Y_n^{(i)} \lbrack & = (m-n)Y_{n+m}^{(i)}, \\
 \lbrack Y_m^{(i)}, Y_n^{(j)} \rbrack  & = 0, \quad &  \lbrack Y_m^{(i)}, J_n^{(jk)} \rbrack & = \left(Y_{n+m}^{(j)} \delta^{ik} - Y_{n+m}^{(k)} \delta^{ij} \right).
\end{aligned}
\end{equation}
where ${f^{ab}}_{c} $ are the structure constants of the Lie algebra $\mathfrak{so}(d)$. 
The two-point functions of this algebra were first given in~\cite{Henkel02,Henkel06b} 
and then re-derived  in~\cite{Bagchi09a, Alishahiha09}. In  the representation (\ref{cga4.4}), they are of the form
$\sim |t|^{-2x} \exp\left[-2{\vec{\gamma}\cdot\vec{r}}/t\right]$. 
Different representations of $\mbox{\sc cga}(1)$ and the resulting
two-point functions are given in \cite{Henkel06b}. 

To obtain representations in $ 1+1 $ dimensions, we first observe that the full \mbox{\sc cga} 
in this dimension can be obtained directly from a contraction of CFT$_2 $. 
To observe this contraction, we go to complex coordinates, $ z=t+\II r/c $. 
Relativistic conformal symmetry of $ d=2 $ contains two copies of the Virasoro algebra:
\begin{equation}
L_n = -z^{n+1} \partial_z, \qquad \bar{L}_n = -\bar{z}^{n+1} \partial_{\bar{z}},
\end{equation}
which are the generators of holomorphic and antiholomorphic transformations. 
Now, we impose the contraction of eq.~(\ref{eq1}), 
and in the limit $ c \to \infty $ we have the generators~\cite{Hosseiny10,Bagchi09a}:
\begin{equation}
X_n = L_n  + \bar{L}_n + \text{O}(1/c), \quad Y_n = - \frac{\II}{c}(L_n - \bar{L}_n)  + \text{O}(1/c).
\end{equation}
A different contraction builds first a different representation of 
$2D$ conformal invariance (which leads to distinct two-point functions)
before contracting \cite{Henkel02,Henkel06b}. 
Now that we have the full \mbox{\sc cga} algebra obtained from contraction of CFT$_2$, 
we might be able to obtain its representations by contraction as well. 
It is not always true that representations of an algebra can be obtained from contraction as well. 
However, in this case it is possible. The result is that the full \mbox{\sc cga} in $d=1$ 
contains to {\em distinct} central charges $c_1$ and $c_2$ \cite{Ovsienko98}:
\begin{equation}
\begin{aligned}
\lbrack X_m, X_n \rbrack  & = (m-n)X_{m+n} + \tfrac{1}{12}c_1 m (m^2 - 1) \delta_{m+n,0}, \\
\lbrack X_m, Y_n \rbrack  & = (m-n)Y_{m+n} + \tfrac{1}{12}c_2 m (m^2 - 1) \delta_{m+n,0}.
\end{aligned}
\end{equation}
The independence of these two central charges $c_{1,2}$ can be seen in a simple way through the
following example: consider the generators $V_n$ and $V_n'$ ($n\in\mathbb{Z}$) of
two commuting Virasoro algebras with central charges $c$ and $c'$. Then
identify
\BEQ
X_n\mapsto \left( \begin{array}{cc} V_n+V_n' & 0\\ 0& V_n+V_n'
\end{array}\right),
\ Y_n\mapsto \left(\begin{array}{cc} 0& V_n\\ 0& 0 \end{array}\right),
\ c_1 \mapsto (c+c')\left(\begin{array}{cc} 1 & 0 \\ 0 & 1 \end{array}\right),
\ c_2  \mapsto c' \left(\begin{array}{cc} 0 & 1 \\ 0 & 0 \end{array}\right)
\EEQ

Two-point functions of the full \mbox{\sc cga} can as well be obtained via contraction:
\begin{equation}
\begin{aligned}
\langle\phi_1(t_1,r_1) \phi_2(t_2,r_2) \rangle_{\text{CGA}} & =  \lim_{c \to \infty} \langle \phi_1(t_1,r_1) \phi_2(t_2,r_2) \rangle_{\text{CFT}}  \\
& =\lim_{c \to \infty} A \delta_{h_1,h_2} \delta_{\bar{h}_1, \bar{h}_2} (t_{12} -  \tfrac{\II}{c}r_{12})^{-(x - \II c \gamma)} (t_{12} -  \tfrac{i}{c}x_{12})^{-(x - \II c \gamma)} \\
& = a \delta_{x_1, x_2} \delta_{\gamma_1, \gamma_2} \, t_{12}^{-2x_1}  \exp \left[ \tfrac{-2\gamma_1 r_{12}}{t_{12}}  \right].
\end{aligned}
\end{equation}
with the abbreviations $t_{12}=t_1-t_2$ and $r_{12}=r-1-r_2$. 
Now, we consider the logarithmic representations and find its contracted form, for the special case $d=1$. 
In the logarithmic representation and taking the simplest rank two Jordan cell, 
we need two states $ \ket{x, \gamma, 0} $ and $ \ket{x, \gamma, 1} $. The action of the generators on the first one
is conventional 
\begin{equation}
\begin{aligned}
X_0 \ket{x, \gamma, 0} &=  x \ket{x, \gamma, 0}, \\
Y_0 \ket{x, \gamma, 0} &=  \gamma \ket{x, \gamma, 0}.
\end{aligned}
\end{equation}
whereas action on the logarithmic partner  $ \ket{x, \gamma, 1} $ is more involved:
\begin{equation}
\begin{aligned}
X_0 \ket{x, \gamma, 1} &=  x \ket{x, \gamma, 1} + \ket{x, \gamma, 0} \\
Y_0 \ket{x, \gamma, 1} &=  \gamma \ket{x, \gamma, 1}.
\end{aligned}
\end{equation}
So, it appears that $ Y_0 $ acts diagonally and logarithms do not involve the rapidity $ \gamma $. 
To derive the logarithmic two-point functions, we follow the contraction approach for logarithmic two-point functions resulting in:
\begin{equation}
\begin{aligned}
\langle  \phi_1(t_1,r_1) \phi_2(t_2,r_2) \rangle_{\text{\sc CGA}} & = 0, \\
\langle \phi_1(t_1,r_1) \psi_2(t_2,r_2)  \rangle_{\text{\sc CGA}} 
& = \lim_{c \to \infty} \langle \phi_1(t_1,r_1) \psi_2(t_2,r_2) \rangle_{\text{CFT}} \\
& = b \delta_{x_1, x_2} \delta_{\gamma_1, \gamma_2} \, t_{12}^{-2x_1}  \exp \left[ -\tfrac{2\gamma_1 r_{12}}{t_{12}}  \right].
\end{aligned}
\end{equation}
For last two-point function again we have: 
\begin{align}
\langle \psi_1(t_1,r_1) \psi_2(t_2,r_2) \rangle_{\text{\sc CGA}} 
& = \lim_{c \to \infty} \langle \psi_1(t_1,r_1) \psi_2(t_2,r_2) \rangle_{\text{CFT}}  \nonumber \\
& = \delta_{x_1, x_2} \delta_{\gamma_1, \gamma_2} \, t_{12}^{-2x_1}  \exp \left[ -\tfrac{2\gamma_1 r_{12}}{t_{12}}  \right] 
\left(d - 2b \log t_{12}\right).
\end{align}
All this procedure may be redone using the nilpotent variables~\cite{Moghimi00}, 
directly deriving the correlators using 
the non-relativistic algebra.\footnote{The extension to $d\geq 1$ dimensions is straightforward \cite{Henkel13b}.}  
We are also able to consider a Jordan cell structure for the rapidity:

\begin{equation}
\begin{aligned}
X_0 \ket{x, \gamma, 1} &=  x \ket{x, \gamma, 1} + x' \ket{x, \gamma, 0} \\
Y_0 \ket{x, \gamma, 1} &=  \gamma \ket{x, \gamma, 1} + \gamma' \ket{x, \gamma, 0}.
\end{aligned}
\end{equation}
And a simple change appears:
\begin{equation}
\langle \psi_1(t_1,r_1) \psi_2(t_2,r_2)  \rangle  =  
\delta_{x_1, x_2} \delta_{\gamma_1, \gamma_2} \, t_{12}^{-2x_1} 
\exp \left[ -2\gamma_1 \tfrac{r_{12}}{t_{12}} \right] \left(-2b x' \log t_{12} - 2a \gamma' \tfrac{r_{12}}{t_{12}} + 2d\right).
\end{equation}
This is an interesting development suggesting that chiral LCFTs should exist.

\subsection{Exotic Galilean Conformal Algebra (\mbox{\sc ecga})}

The algebra $\mbox{\sc cga}(2)$ admits an extra central charge in $ (2+1) $ dimensions~\cite{Lukierski06}. 
In fact the commutator of boosts is no longer vanishing, reminiscent of non-commutative theories:
\begin{equation}
\lbrack B_i, B_j \rbrack = \mathcal{B} \epsilon_{ij}, \qquad \lbrack P_i, K_j \rbrack = -2\mathcal{B} \epsilon_{ij},
\end{equation}
where $ \epsilon_{ij} $ is the antisymmetric $2$-dimensional tensor. Here $ \mathcal{B} $ commutes with all generators 
of the algebra and is therefore an extra central charge. 
Its physical significance has been of interest~\cite{Duval09}. 
The central charge can also be obtained by contraction and two-point function is realised using auxiliary coordinates~\cite{Martelli09}.
Examples of systems of non-linear equations with the {\sc ecga} as a dynamical symmetry are given in \cite{Cherniha10}. 

To proceed we follow~\cite{Martelli09} and introduce an auxiliary internal space with three dimensions 
$ w $, $ v_1 $, $ v_2 $ (therefore we have a six-dimensional space now). 
Using these coordinates new differential operators for the generators of \mbox{\sc ecga} may be written:
\begin{equation}
\begin{aligned}
 H  & = - \partial_t,  & D  & = -r_i \partial_i - t \partial_t - x,  &  K & =  -2t r_i\partial_i - t^2 \partial_t - 2r_i \chi_i, \\
P_i & = - \partial_i, & B_i & = -t\partial_i - \chi_i,  & K_i &=  -t^2 \partial_i + 2t \chi_i - 2r_j \epsilon_{ij} \gamma \\
J & = -\epsilon_{ij}r_j \partial_j - \tfrac{1}{2\gamma} \chi_i \chi_i,  
& \mathcal{B} & = \partial_{w},  &  \chi_i & =  \partial_{v_i} - \tfrac{1}{2}\epsilon_{ij}v_j \partial_{w}.
\end{aligned}
\end{equation}
The operator $ J $ plays the role of rapidity here. 
In this realisation one desires local operators to be simultaneous eigenstates of $ D $ and $ \mathcal{B} $:
\begin{equation}
 \lbrack D,\phi \rbrack = x \phi, \qquad  \lbrack \mathcal{B}, \phi \rbrack = \gamma \phi.
\end{equation}
Now, if we look for the most general case, local fields will have to be eigenstates of $ K_i $ as well. 
The two point function of \mbox{\sc ecga} (without rapidity) has been worked out~\cite{Martelli09}:
\begin{equation}
\langle \phi_1(t_1,\boldsymbol{r_1}) \phi_2(t_2,\boldsymbol{r_2}) \rangle   =  \delta_{x_1, x_2} \delta_{\gamma_1+ \gamma_2,0} \, 
t^{-2x_1} \exp \left[\tfrac{1}{2}\gamma \epsilon_{ij}(\tfrac{1}{4}v_j^+- u_j)v_i \right] \,
 \mathcal{O}_1 (u_i + \tfrac{1}{2}v_i^+ ),
\end{equation}
in which $ \mathcal{O} $ is an arbitrary function, $ \boldsymbol{u} = (\boldsymbol{r}_1-\boldsymbol{r}_2)/(t_1-t_2 ) $ and $ v^+=  v_1 + v_2 $. 
Here, we add the explicit dependence on the \mbox{\sc ecga}-rapidity 
\begin{align}
 \langle \Phi_1( t_1,\boldsymbol{r}_1, \boldsymbol{v}_1) & \Phi_2( t_2,\boldsymbol{r}_2, \boldsymbol{v}_2)  \rangle = \nonumber \\
 & t^{-x^+} \exp \left[ -\tfrac{1}{2} (\lambda_i - \tfrac{1}{4}\gamma \epsilon_{ij} v_j^+  + \gamma \epsilon_{ij}u_j)v_i  -u_i\lambda_i^+
 \right]   \, \mathcal{O}_1( \boldsymbol{u} - \tfrac{1}{2}\boldsymbol{v}^+ ),
\end{align}
where $ \lambda^+ = \lambda_1 + \lambda_2 $ and $ x^+ = x_1 + x_2 $. The function $ \mathcal{O}_1 $ now satisfies some constraints~\cite{Hosseiny:2013ex}. To find the logarithmic version of the two point functions one can add a nilpotent variable to the fields and after some algebra one finds:
\begin{equation}
\begin{aligned}
\langle \Phi_1( t_1,\boldsymbol{r}_1, \boldsymbol{v}_1) \Phi_2(t_2,\boldsymbol{r}_2, \boldsymbol{v}_2) \rangle & = 0, \\
\langle \Phi_1( t_1,\boldsymbol{r}_1, \boldsymbol{v}_1) \Psi_2(t_2,\boldsymbol{r}_2, \boldsymbol{v}_2)  \rangle & 
= t^{-x^+} \exp \left[ -\tfrac{1}{2} (\lambda_i - \tfrac{1}{4}\gamma \epsilon_{ij} v_j^+  + \gamma \epsilon_{ij}u_j)v_i - u_i\lambda_i^+ \right] \delta_{x,0} \delta_{\gamma^+,0}   \\
& \times \mathcal{O}_2( \boldsymbol{u} - \tfrac{1}{2}\boldsymbol{v}^+ ), \\
\langle \Psi_1( t_1,\boldsymbol{r}_1, \boldsymbol{v}_1) \Psi_2(t_2,\boldsymbol{x}_2, \boldsymbol{v}_2) \rangle & = 
t^{-x^+} \exp \left[ -\tfrac{1}{2} (\lambda_i - \tfrac{1}{4}\gamma \epsilon_{ij} v_j^+  + \gamma \epsilon_{ij}u_j)v_i - u_i\lambda_i^+ \right] \delta_{x,0} \delta_{\gamma^+,0} \\
& \times \left[    -2\mathcal{O}_1(\boldsymbol{u} - \tfrac{1}{2}\boldsymbol{v}^+ ) -  2(u_i \lambda'_i + x' \ln t) \, \mathcal{O}_2(\boldsymbol{u} - \tfrac{1}{2}\boldsymbol{v}^+) \right].
\end{aligned}
\end{equation}
We now have two arbitrary functions $\mathcal{O}_{1,2}$ involved. 

In view of possible relationships with non-equilibrium statistical physics, one may ask whether causality conditions might be
derived analogously to the $\wit{\mathfrak{sch}}(d)$ algebra, see section~3. Indeed, for the {\sc ecga} the extra central generator
might provide the basis for an extension to a new parabolic sub-algebra, which likely could turn out to be $B_3$. This is an open
problem to which we hope to return in the future.


\section{Logarithmic extension of the ageing algebra $\mathfrak{age}(d)$}

\subsection{Physical ageing}

A paradigmatic example of cooperative non-equilibrium dynamics are {\em ageing phenomena}, 
see e.g. \cite{Bray94a,Cugliandolo02,Henkel10} for introductions and reviews. These occurs for instance
if the temperature $T$ of a system, initial prepared in a disordered initial state, is quenched to some value $T\leq T_c$ below or at
the critical temperature $T_c>0$. The quench brings the system out of equilibrium and the long-time relaxation dynamics typically
displays dynamical scaling, even if the stationary state itself need not be critical. In this paper, we shall concentrate on a
single aspect, namely the dynamical scaling of the (linear) 
auto-response of the order parameter $\phi(t,\vec{r})$ with respect to a perturbation
in its canonically conjugate field $h(s,\vec{r})$: 
\BEQ \label{5.1}
R(t,s) := \left.\frac{\partial \langle \phi(t,\vec{r})\rangle}{\partial h(s,\vec{r})}\right|_{h=0} 
= \left\langle \phi(t,\vec{r}) \wit{\phi}(s,\vec{r})\right\rangle = s^{-1-a} f_R\left(\frac{t}{s}\right)
\EEQ
The scaling form is valid in the double scaling limit $t,s\to\infty$ with $y=t/s>1$ fixed (this implies that one must have
$t-s\to\infty$). If $y\gg 1$, one generically expects $f_R(y)\sim y^{-\lambda_R/z}$. Finally, re-writing $R(t,s)$ as a correlator
between the order-parameter $\phi$ and an associated {\em `response field'} 
$\wit{\phi}$ is a well-known consequence of Janssen-de Dominicis theory \cite{Janssen92,Cugliandolo02} and we shall use this
below for the derivation of explicit expressions of the scaling function $f_R(y)$. 

\subsection{Generators}

Physical ageing occurs far from equilibrium and time-translation-invariance does not hold. 
Only sub-algebras of $\mathfrak{sch}(d)$ without time-translations can therefore be candidates for dynamical symmetries of ageing. 
Here, we consider the {\em ageing algebra} 
$\mathfrak{age}(d) := 
\langle X_{0,1},Y_{\pm 1/2}^{(j)}, M_0, R_0^{(jk)}\rangle_{j,k=1,\ldots,d}\subset \mathfrak{sch}(d)$, 
which is a sub-algebra of the Schr\"odinger algebra, see figure~\ref{fig1ab}a. The embedding $\mathfrak{age}(1)\subset B_2$ and
the parabolic extension is illustrated in figure~\ref{fig2}. 

With respect to the generators of the representation (\ref{2.3}), it turns out that $\mathfrak{age}(d)$ admits more general representations
which contain a second, new scaling dimension $\xi$ \cite{Picone04,Henkel06a}\footnote{If one assumes time-translation-invariance, 
the commutator $[X_1,X_{-1}]=2X_0$ leads to $\xi=0$. A well-known exactly solved example with $\xi\ne 0$ is the $1D$ Ising model
with Glauber dynamics, quenched to $T=0$. Further examples will be discussed below.}
\BEQ \label{5.2}
X_n = -t^{n+1}\partial_t - \frac{n+1}{2}t^n \vec{r}\cdot\vec{\nabla}_{\vec{r}} 
- \frac{\cal M}{2}(n+1)n t^{n-1} \vec{r}^2
- \frac{n+1}{2} x t^n - (n+1)n \xi t^n 
\EEQ
where now $n\geq 0$. The other generators are still given by (\ref{2.3}) and the commutators (\ref{gl:5:svComm}) 
remain valid. Still, this does {\em not} exhaust the possible representations of $\mathfrak{age}(d)$ \cite{Minic12}, but since the
relationship with logarithmic scaling has not yet been explored, we shall not discuss these here.  
For the derivation of auto-responses, it is enough to concentrate on the temporal part 
$\langle \Psi(t_1,\vec{r}_0)\Psi(t_2,\vec{r}_0)\rangle$, 
the form of which is described by the two generators $X_{0,1}$, with
the commutator $[X_1,X_0]=X_1$.  

Logarithmic representation of $\mathfrak{age}(d)$, 
analogously to section~2, can be constructed by replacing {\em both} scaling dimensions $x$ and $\xi$ by matrices \cite{Henkel13b}
\BEQ \label{5.3}
x \mapsto \left(\matz{x}{x'}{0}{x}\right) \;\; , \;\;
\xi  \mapsto \left(\matz{\xi}{\xi'}{\xi''}{\xi}\right)
\EEQ
in eq.~(\ref{5.2}). The other generators (\ref{2.3}) are kept unchanged. 
Without restriction of generality, one can always achieve either a diagonal form (with $x'=0$) 
or a Jordan form (with $x'=1$) of the first matrix, but the structure of the second matrix in 
(\ref{5.3}) has to be clarified. Setting
$\vec{r}=\vec{0}$, we have from (\ref{5.2}) the two generators
\BEQ \label{5.4}
X_0 = - t\partial_t - \demi \left(\matz{x}{x'}{0}{x}\right) \;\; , \;\;
X_1 = - t^2\partial_t - t \left(\matz{x+\xi}{x'+\xi'}{\xi''}{x+\xi}\right) 
\EEQ
and $[X_1,X_0]=X_1 +\demi t \,x'\xi'' \left(\matz{-1}{0}{0}{1}\right) \stackrel{!}{=} X_1$. 
Hence $x' \xi''\stackrel{!}{=}0$ and one must distinguish two cases.
\begin{enumerate}
\item $x'=0$. The first matrix in (\ref{5.3}) is diagonal. 
In this situation, there are two distinct possibilities: (i) either, the matrix
$\left(\matz{\xi}{\xi'}{\xi''}{\xi}\right)\rar\left(\matz{\xi_+}{0}{0}{\xi_-}\right)$ 
is diagonalisable. One then has
a pair of quasi-primary operators, with scaling dimensions $(x,\xi_+)$ and $(x,\xi_-)$. 
This reduces to the standard
form of non-logarithmic local scale-invariance \cite{Henkel06a}.  Or else, (ii), the matrix 
$\left(\matz{\xi}{\xi'}{\xi''}{\xi}\right)\rar\left(\matz{\bar{\xi}}{1}{0}{\bar{\xi}}\right)$ 
reduces  to a Jordan form.
This is a special case of the situation considered below.
\item $\xi''=0$. Both matrices in (\ref{2.2}) 
reduce simultaneously to a Jordan form. While one can always normalise such 
that either $x'=1$ or else $x'=0$, there is no obvious normalisation for $\xi'$. 
This is the main case which we shall study below.
\end{enumerate}
In conclusion: {\em without restriction on the generality, 
one can set $\xi''=0$ in eqs.~(\ref{5.3},\ref{5.4}).} 

We point out that the scaling dimension $\xi$ is identical to the one used in the parabolic sub-algebra $\wit{\mathfrak{sch}}(1)$ in
eq.~(\ref{3.3}) in section~3.  Therefore, an extension to the parabolic sub-algebra $\wit{\mathfrak{age}}(1)$ will produce
the analogous causality constraints on the two-point functions, when then should be interpreted as responses, 
and not as correlators \cite{Henkel12a,Henkel13b}.  

\subsection{Two-point functions}

Consider the following two-point functions, built from the components of quasi-primary operators
of logarithmic {\sc lsi}
\BEA
F = F(t_1, t_2) &:=& \left\langle \phi_1(t_1)\phi_2(t_2)\right\rangle \nonumber \\
G_{12} = G_{12}(t_1, t_2) &:=& \left\langle \phi_1(t_1)\psi_2(t_2)\right\rangle \nonumber \\
G_{21} = G_{21}(t_1, t_2) &:=& \left\langle \psi_1(t_1)\phi_2(t_2)\right\rangle  \\
H = H(t_1, t_2) &:=& \left\langle \psi_1(t_1)\psi_2(t_2)\right\rangle \nonumber 
\EEA
Their co-variance under the representation (\ref{2.3}), with $\xi''=0$, 
leads to a system of eight linear equations for a set of four functions in two variables. 
There is an unique solution, up to normalisations. It reads \cite{Henkel13b} 
\BEA
F(t_1,t_2) &=& t_2^{-(x_1+x_2)/2}\: y^{\xi_2 +(x_2-x_1)/2} (y-1)^{-(x_1+x_2)/2-\xi_1-\xi_2} f_0 
\nonumber \\
G_{12}(t_1,t_2) &=& t_2^{-(x_1+x_2)/2}\: y^{\xi_2 +(x_2-x_1)/2} (y-1)^{-(x_1+x_2)/2-\xi_1-\xi_2} 
\Bigl( g_{12}(y) + \ln t_2 \cdot \gamma_{12}(y) \Bigr) 
\nonumber \\
G_{21}(t_1,t_2) &=& t_2^{-(x_1+x_2)/2}\: y^{\xi_2 +(x_2-x_1)/2} (y-1)^{-(x_1+x_2)/2-\xi_1-\xi_2} 
\Bigl( g_{21}(y) + \ln t_2 \cdot \gamma_{21}(y) \Bigr)   
\nonumber \\
H(t_1,t_2) &=& t_2^{-(x_1+x_2)/2} \: y^{\xi_2 +(x_2-x_1)/2} (y-1)^{-(x_1+x_2)/2-\xi_1-\xi_2} 
\label{3.17} \\
& & \times \Bigl( h_0(y) + \ln t_2 \cdot h_1(y) +\ln^2 t_2 \cdot h_2(y) \Bigr) \nonumber
\EEA
where the scaling functions, depending only on $y=t_1/t_2$, are given by 
\BEA
g_{12}(y) &=& g_{12,0} +\left(\frac{x_2'}{2}+\xi_2'\right) f_0 \ln \left|\frac{y}{y-1}\right| 
\nonumber \\
g_{21}(y) &=& g_{21,0} -\left(\frac{x_1'}{2}+\xi_1'\right) f_0 \ln |y-1| - \frac{x_1'}{2} f_0 \ln |y| 
\nonumber \\
h_0(y) &=& h_0 - \left[ \left(\frac{x_1'}{2}+\xi_1'\right)g_{21,0} + 
\left(\frac{x_2'}{2}+\xi_2'\right)g_{12,0}\right]\ln|y-1| 
- \left[ \frac{x_1'}{2} g_{21,0} - \left(\frac{x_2'}{2}+\xi_2'\right)g_{12,0}\right]\ln|y| 
\nonumber \\
& &  + \demi f_0 \left[ \left( \left(\frac{x_1'}{2} +\xi_1'\right)\ln |y-1| 
+ \frac{x_1'}{2}\ln |y|\right)^2 
-  \left(\frac{x_2'}{2} +\xi_2'\right)^2 \ln^2\left|\frac{y}{y-1}\right| \right]  \label{3.16}
\EEA
where 
\BEA 
\gamma_{12}(y) &=& -\demi x_2' f_0 \;\;\;\; , \;\;\;\; \gamma_{21}(y) = - \demi x_1' f_0 \nonumber \\ 
\label{3.14}
h_1(y) &=& - \demi\left( x_1' g_{12}(y) + x_2' g_{21}(y) \right) \;\; , \;\;
h_2(y) = \frac{1}{4} x_1' x_2' f_0
\EEA
and $f_0, g_{12,0}, g_{21,0}, h_0$ are normalisation constants.

The solution $F(t_1,t_2)$ does not vanish, in contrast to logarithmic Schr\"odinger- or logarithmic conformal Galilean invariance. 
Rather, it leads to the scaling function of non-logarithmic local scale-invariance ({\sc lsi}) of the autoresponse, cf (\ref{5.1})
and including the causality condition $y>1$
\BEQ \label{R}
f_R(y) = f_0\, y^{1+a'-\lambda_R/z} (y-1)^{-1-a'} \Theta(y-1)
\EEQ
where the ageing exponents $a,a',\lambda_R$ are related to the scaling dimensions as follows:
\BEQ
a =\demi(x_1+x_2) -1 \;\; , \;\; a'-a = \xi_1 + \xi_2 \;\; , \;\; \lambda_R = 2 (x_1 +\xi_1)
\EEQ
For example, the exactly solvable $1D$ kinetic Ising model with 
Glauber dynamics at zero temperature \cite{Godreche00a} 
satisfies (\ref{R}) with the values $a=0, a'-a=-\demi, \lambda_R=1, z=2$ \cite{Picone04}.

Although the algebra $\mathfrak{age}(d)$ was written down for a dynamic exponent $z=2$, 
the form of the auto-responses  is essentially independent of this feature. The change 
$(x,x',\xi,\xi')\mapsto \bigl((2/z) x, (2/z) x', (2/z) \xi, (2/z)\xi'\bigr)$ 
gives the form valid for an arbitrary dynamical exponent $z$.

Comparison with the results 
of logarithmic Schr\"odinger- or conformal Galilean-invariance shows:
\begin{enumerate}
\item Logarithmic contributions may arise, either as corrections 
to the scaling behaviour via additional powers of $\ln t_2$, or else
through logarithmic terms in the scaling functions themselves. 
These can be described independently in terms of the parameter sets $(x_1',x_2')$ and
$(\xi_1',\xi_2')$. 

In particular, it is possible to have representations of 
$\mathfrak{age}(d)$ with an explicit doublet in only one of the
two generators $X_0$ and $X_1$. 
\item Logarithmic corrections to scaling arise if either $x_1'\ne0$ or $x_2'\ne 0$, 
but the absence of time-translation-invariance
allows for the presence of quadratic terms in $\ln t_2$. 
\item If one sets $x_1'=x_2'=0$, 
there is no breaking of dynamical scaling through logarithmic corrections. 
However, the scaling functions 
$g_{12}(y), g_{21}(y)$ and $h_0(y)$ may still contain logarithmic terms. 

This is qualitatively distinct from logarithmic Schr\"odinger-invariance (\ref{2.29}): for example 
$H(y t,t) = 
\delta_{x_1,x_2}\,t^{-x_1} \left( H_0 - G_0 \ln (y-1) - G_0 \ln t_2 \right) (y-1)^{-x_1}$, 
such that logarithmic corrections to scaling, parametrised by $G_0$, 
are coupled to a corresponding term in the scaling function itself.  
\item If time-translation-invariance is assumed, one has 
$\xi_1=\xi_2=\xi_1'=\xi_2'=0$, $x_1=x_2$ and $f_0=0$ 
and one is back to logarithmic Schr\"odinger-invariance (\ref{2.29}).  
\end{enumerate}

\section{Applications} 

We now briefly review discuss two candidate models for an application of logarithmic {\sc lsi} 
({\sc llsi}) in physical ageing \cite{Henkel13b}. The universality classes of both the Kardar-Parisi-Zhang equation 
and directed percolation are widely considered to
be the most simple models for the non-equilibrium phase transitions they describe. It is now well-established
that they both undergo ageing in the sense that the three defining properties listed in the introduction are satisfied, see
e.g. \cite{Kall99,Dornic01,Enss04,Ramasco04,Daqu11,Henk12,Hyun12}.  

\subsection{One-dimensional Kardar-Parisi-Zhang equation}

When describing the growth of interfaces, a lattice model can be formulated in terms
of time-dependent heights $h_i(t)\in\mathbb{N}$ (and $i\in\mathbb{Z}$), 
and subject to a stochastic deposition of particles. 
If one further admits a RSOS constraint of the form 
$0\leq |h_{i+1}(t)-h_i(t)|\leq 1$ \cite{Kim89}, this
goes in a continuum limit to the paradigmatic model 
equation proposed by Kardar, Parisi and Zhang (KPZ) 
\cite{Kard86}, described by a
time-dependent height variable $h=h(t,{r})$
\BEQ \label{kpz}
\frac{\partial h}{\partial t} = \nu \frac{\partial^2 h}{\partial {r}^2} 
+ \frac{\mu}{2} \left( \frac{\partial h}{\partial {r}}\right)^2 +\eta
\EEQ
where $\eta(t,{r})$ is a white noise with zero mean and variance 
$\langle\eta(t,{r})\eta(t',{r}')\rangle=2\nu T\delta(t-t') \delta({r}-{r}')$ 
and $\mu,\nu,T$ are material-dependent constants. 
In $1D$ the height distribution can be shown to converge for large times 
towards the gaussian Tracy-Widom distribution \cite{Sasa10,Cala11,Geudre12}. 
The numerous application of KPZ include Burgers turbulence, 
directed polymers in a random medium, glasses and
vortex lines, domain walls and biophysics, 
see e.g. \cite{Bara95,Halp95,Krug97,Krie10,Sasa10b,Tong94,Batc00,Corwin11}
for reviews. 
Experiments on the growing interfaces of turbulent liquid crystals 
reproduce this universality class \cite{Take11}. 

Since the main prediction of {\sc llsi} concerns the response, we focus exclusively on this. Indeed, by varying the
deposition rate of particles onto the surface, up to a waiting time $s$, one may numerically find the time-integrated autoresponse 
\BEQ \label{4.2}
\chi(t,s) = \int_0^s \!\!\D u\: R(t,u) = 
\frac{1}{L} \sum_{i=1}^L \left\langle \frac{h_{i}^{(A)}(t;s) -
h_{i}^{(B)}(t)}{\eps a_i}\right\rangle  
= s^{-a} f_{\chi}\left( \frac{t}{s}\right)
\EEQ
together with the generalised Family-Vicsek scaling \cite{Kall99,Bust07,Chou10,Daqu11,Krec97,Henk12}. 
The autoresponse exponent is read off from 
$f_{\chi}(y)\sim y^{-\lambda_R/z}$ for $y\to\infty$. In $1D$, one has the well-known exponents
$a=-1/3$, $\lambda_R=1$ and $z=3/2$. 

\begin{figure}[tb]
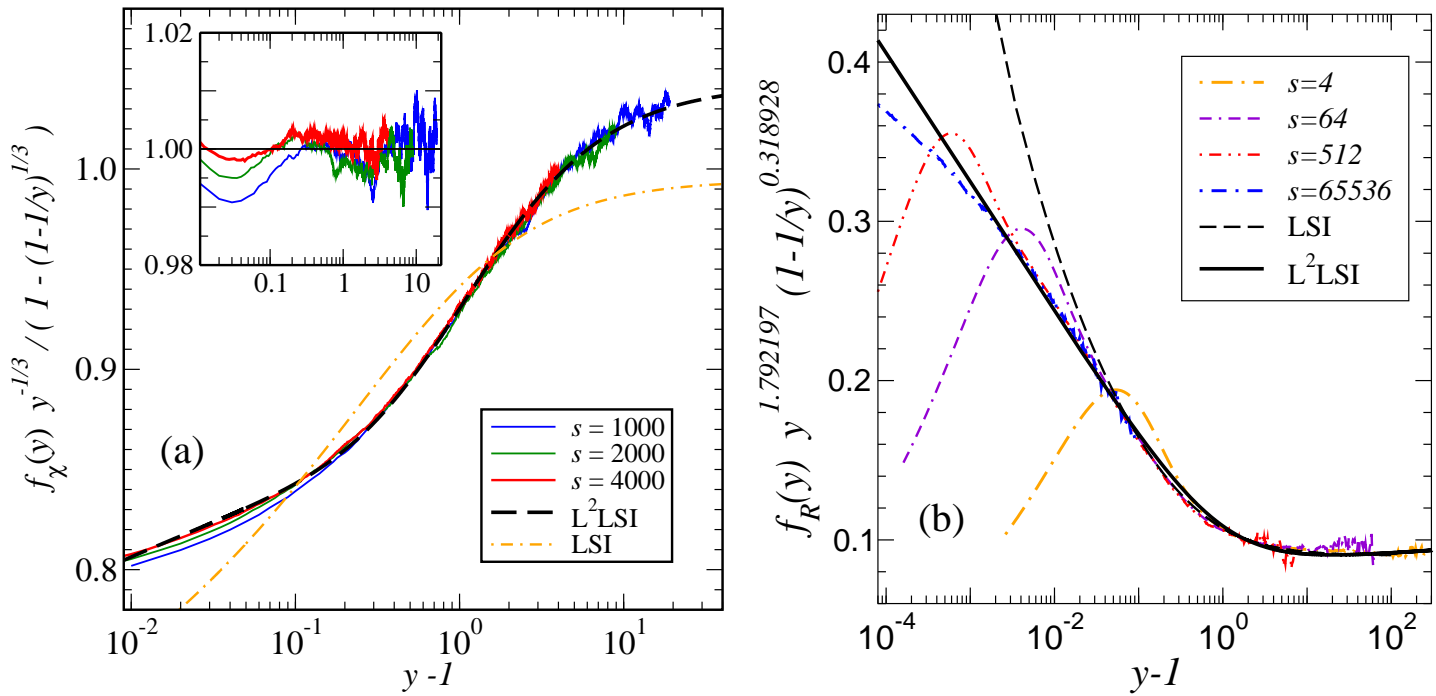

\centerline{\psfig{figure=log_sch_age_fig3a.eps,width=3.75in,clip=} ~ \psfig{figure=log_sch_age_fig3b.eps,width=3.55in,clip=}}
\caption[fig3]{\label{fig3} (a) Scaling of the integrated autoresponse $\chi(t,s) = s^{+1/3} f_{\chi}(t/s)$ of the $1D$
Kardar-Parisi-Zhang equation, as a function of $y=t/s$, for several values of the waiting time $s$.  
The figure shows the reduced scaling function  
$f_{\rm red}(y) = f_{\chi}(y) y^{-1/3} \left[1-(1-y^{-1})^{1/3}\right]^{-1}$.  
The dash-dotted line labelled {\sc lsi} gives a fit to non-logarithmic {\sc lsi} (see text) and the 
dashed line labelled {\sc l$^2$lsi} gives the prediction (\ref{4.5}). 
The inset shows the ratio $f_{\chi}(y)/f_{\rm L^2LSI}(y)$ over against $y$. \\
(b) Reduced scaling function $h_R(y) = f_R(y) y^{\lambda_R/z} (1-1/y)^{1+a}$ 
the autoresponse $R(t,s)=s^{-1-a}f_R(t/s)$ 
of the $1D$ critical contact process, as a function of $y=t/s$, for several
values  of the waiting time $s$.  
The dashed line labelled `{\sc lsi}' is from (\ref{R}), with $a'-a=0.26$. The full curve labelled
`{\sc l$^2$lsi}' is obtained from eq.~(\ref{4.8}), derived from logarithmic {\sc lsi} with $f_0=0$, see text.
After \cite{Henk12,Henkel13b}.}
\end{figure}

Following \cite{Henk12}, 
in order to compare the data in figure~\ref{fig3}a with the prediction (\ref{3.17}) (with the tacit extension to generic $z$
mentioned above), we first make the working hypothesis that $R(t,s) = \langle \psi(t) \wit{\psi}(s)\rangle$,  
where the two scaling operators $\psi$ and $\wit{\psi}$ 
are described by the logarithmically extended scaling dimensions
\BEQ
\left(\matz{x}{x'}{0}{x}\right) \;\; , \;\; 
\left(\matz{\xi}{\xi'}{0}{\xi}\right) \;\; \mbox{\rm ~~and~~} \;\; 
\left(\matz{\wit{x}}{\wit{x}'}{0}{\wit{x}}\right) \;\; , \;\; 
\left(\matz{\wit{\xi}}{\wit{\xi}'}{0}{\wit{\xi}}\right) 
\EEQ 
In view to good quality of the data collapse, we assume that logarithmic corrections to scaling should be absent, hence
$x'=\wit{x}'=0$ in view of (\ref{3.14}). In addition, the requirement of a simple power-law form for $y\gg 1$ leads to
$\xi'=0$ and one can then normalise $\wit{\xi}'=1$. 
With the scaling form $R(t,s) = \left\langle \psi(t)\wit{\psi}(s)\right\rangle=s^{-1-a} f_R(t/s)$, it remains 
\BEQ \label{4.4}
f_R(y) = y^{-\lambda_R/z} \left( 1 - y^{-1}\right)^{-1-a'} 
\left[ h_0 - g_0 \ln\left( 1 - y^{-1}\right) - \demi f_0  \ln^2\left( 1 - y^{-1}\right)\right] 
\EEQ
with the exponents $1+a = (x + \wit{x})/z$, $a'-a = \frac{2}{z} \left( \xi + \wit{\xi}\,\right)$, 
$\lambda_R/z = x + \xi$ and the normalisation constants $h_0, g_0=g_{12,0},f_0$. 
Using the specific value $\lambda_R/z-a=1$ which holds for the $1D$ KPZ, the integrated autoresponse 
$\chi(t,s) = s^{-a} f_{\chi}(t/s)$ becomes  
\BEQ \label{4.5} 
f_{\chi}(y) = y^{+1/3} 
\left\{ A_0 \left[ 1 - \left( 1- y^{-1}\right)^{-a'} \right] \right.  
+ \left.  \left( 1 - y^{-1}\right)^{-a'} \left[ A_1 \ln\left( 1 - y^{-1}\right) 
+ A_2  \ln^2\left( 1 - y^{-1}\right) \right] \right\}
\EEQ
where $A_{0,1,2}$ are normalisations related to $f_0,g_0,h_0$. Indeed, for $y\gg 1$, 
one has $f_{\chi}(y) \sim y^{-2/3}$, as expected. 
The non-logarithmic case would be recovered for $A_1=A_2=0$. 

In figure~\ref{fig3}a, the simulational data from \cite{Henk12}  
are compared with the predicted form (\ref{4.5}). Very large values of the waiting time $s$ are required,
and one may observe from  figure~\ref{fig3}a that even data with $s<10^3$ are not yet fully in the scaling regime. 
While non-logarithmic {\sc lsi} gives an overall agreement with an accuracy up to about $5\%$, when
$a'=-0.5$ is assumed, clear and systematic deviations remain. It turns out that if one tried to use the
more restricted prediction (\ref{2.29}) of logarithmic Schr\"odinger-invariance (without the second scaling dimension), the numerical
result is indistinguishable from non-logarithmic {\sc lsi} \cite{Henkel13b}. However, the prediction (\ref{4.5}) 
reproduces the data to an accuracy better than $0.1\%$ and at least down to $y=t/s\approx 1.03$ (the inset shows that this 
is about the region where the numerical data obey dynamical scaling), with the fitted values
$a'=-0.8206$, $A_0 = 0.7187$, $A_1 = 0.2424$ and $A_2 = -0.09087$.

\subsection{One-dimensional critical directed percolation} 

The directed percolation universality class is usually considered to most simple example of a non-equilibrium phase transition
with an absorbing state. It has been realised in countless different ways, 
with often-used examples being either the contact process or else Reggeon field theory, 
and very precise estimates of the location of the critical point and the critical exponents are known, 
see \cite{Hinrichsen00,Odor04,Henkel09} and references therein. 
Its predictions are also in agreement with extensive recent experiments in turbulent liquid crystals \cite{Takeuchi09}.  
Since it is well-understood that critical $2D$ {\em isotropic}
percolation can be described in terms of conformal invariance 
\cite{Langlands94},\footnote{Cardy \cite{Cardy92} 
and Watts \cite{Watts96} used conformal invariance to derive their celebrate formul{\ae} 
for the crossing probabilities. A precise formulation of the conformal invariance methods 
required in their derivations actually leads to a 
logarithmic conformal field theory \cite{Mathieu07}.} one might wonder whether
some kind of local scale-invariance might be applied to directed percolation.

In the contact process, a response function can be defined by considering the 
response of the time-dependent particle concentration with respect to a time-dependent 
particle-production rate. In figure~\ref{fig3}b, numerical data for the rescaled scaling function 
\BEQ
h_R(y) := f_R(y) y^{\lambda_R/z} (1-y^{-1})^{1+a}
\EEQ
are shown, where the values of the exponents are taken from \cite{Henkel09}. An excellent data collapse is seen. 
Non-logarithmic {\sc lsi}, and assuming $a'-a=0.26$, describes the data well down to about $y\approx 1.1$, but systematic
deviations remain. 

In order to compare the data with logarithmic {\sc lsi}, we make the same working assumptions as before for KPZ. With
$x'=\wit{x}'=0$, logarithmic {\sc lsi} eq.~(\ref{3.16}) predicts
\BEA
h_R(y) &=&  \left( 1 - \frac{1}{y}\right)^{a-a'} 
\left( h_0 - g_{12,0} \wit{\xi}'  \ln (1-1/y) - \demi f_0 \wit{\xi}'^2 \ln^2 (1-1/y)
\right. \nonumber \\
& & \left. ~~~~~~- g_{21,0} \xi'  \ln (y-1) + \demi f_0 \xi'^2 \ln^2 (y-1) \right) 
\label{4.8}
\EEA
Further constraints must be obeyed, in particular 
the resulting scaling function should always be positive. 

Numerical experiments reveal that the best fits are 
obtained by fitting the generic form (\ref{4.8}) to the data. It then
turns out that the terms which depend quadratically 
on the logarithms have amplitudes which are about $10^{-4}$ times smaller
than those of the other terms. We consider this as evidence that $f_0=0$. 
This gives the phenomenological scaling form 
$h_R(y)= h_0 (1-1/y)^{a-a'} \left( 1 - (A+B)\ln(1-1/y) +B\ln (y-1)\right)$,
where $h_0$ is a normalisation constant and 
$A,B$ are two positive universal parameters. 
With the fitted parameters $a-a'=0.00198$, $A=0.407$, $B=0.02$ and $h_0 = 0.08379$, this
gives a good description of the data, down to 
$y-1\approx 2 \cdot 10^{-3}$.  (for smaller values of $y$, we cannot be sure to be still
in the scaling regime). 

Note that the estimate $a'-a\simeq -0.002$ 
is quite distinct from the earlier estimate $a'=a\approx 0.27$ \cite{Henkel06a}
and also implies a small logarithmic contribution in the $y\gg 1$ limit. 

Similar results have also been obtained for the critical $2D$ voter model on a triangular lattice. This 
will be reported elsewhere \cite{Henkel13}.

\section{Conclusions}

We have presented current ideas on the analogues of logarithmic conformal invariance in non-relativistic
contexts. Several formal developments can be carried out in quite close analogy with the well-known conformal
case, but the possibility of true projective representations and the absence of time-translation-invariance in several
physical applications leads to new features, absent from conformal invariance. 

Up to date, the only physical applications studied involve slow relaxation phenomena far from equilibrium. 
It appears that {\em non-equilibrium} scaling operators are to be described in terms of {\em two}, rather than one,
independent scaling exponents, which we labeled here $x$ and $\xi$; and furthermore, in the known physical examples
it seem that only the elusive second scaling dimension $\xi$ is extended to a Jordan matrix and thus carries
the essential logarithmic structure. Further work will without doubt inform us in the future to what extent this
appreciation will remain valid. 

In any case, in very commonly studied models of non-equilibrium statistical physics, NRLCFTs have found their first applications. 

\noindent 
{\bf Acknowledgement:} We thank the organisers of the meeting ``Advanced conformal field theory'' 
at the Institut Poincar\'e in Paris for their warm hospitality. 
MH was partly supported by the Coll\`ege Doctoral franco-allemand Nancy-Leipzig-Coventry
(Syst\`emes complexes \`a l'\`equilibre et hors \`equilibre) of UFA-DFH. 



\end{document}